\documentclass[aip,apm,preprint]{revtex4-1}

\usepackage{graphicx}
\usepackage{epstopdf}
\usepackage[ansinew]{inputenc}
\usepackage{array}
\usepackage{color}
\usepackage{amsmath}
\usepackage{amsxtra}
\usepackage{amstext}
\usepackage{amssymb}
\usepackage{latexsym}
\usepackage{dsfont}
\usepackage{soul}

\begin{document}
\title{Band Structure Mapping of Bilayer Graphene via Quasiparticle Scattering}
\author{Matthew Yankowitz}
\affiliation{Physics Department, University of Arizona, 1118 E 4th Street, Tucson, AZ 85721, USA}
\author{Joel I-Jan Wang}
\affiliation{Department of Physics, Massachusetts Institute of Technology, Cambridge, MA 02138, USA}
\affiliation{School of Engineering and Applied Sciences, Harvard University, Cambridge, MA 02138, USA}
\author{Suchun Li}
\affiliation{Department of Physics, Faculty of Science, National University of Singapore, 2 Science Drive 3, Singapore 117551}
\affiliation{Institute of High Performance Computing, Agency for Science, Technology and Research, 1 Fusionopolis Way, \#16-16 Connexis, Singapore 138632}
\affiliation{NUS Graduate School for Integrative Sciences and Engineering, National University of Singapore, 28 Medical Drive, Singapore 117456}
\author{A. Glen Birdwell}
\affiliation{Sensors and Electron Devices Directorate, US Army Research Laboratory, Adelphi, MD 20783, USA}
\author{Yu-An Chen}
\affiliation{Department of Physics, Massachusetts Institute of Technology, Cambridge, MA 02138, USA}
\author{Kenji Watanabe}
\author{Takashi Taniguchi}
\affiliation{National Institute for Materials Science, 1-1 Namiki, Tsukuba 305-0044, Japan}
\author{Su Ying Quek}
\affiliation{Department of Physics, Faculty of Science, National University of Singapore, 2 Science Drive 3, Singapore 117551}
\affiliation{Institute of High Performance Computing, Agency for Science, Technology and Research, 1 Fusionopolis Way, \#16-16 Connexis, Singapore 138632}
\author{Pablo Jarillo-Herrero}
\affiliation{Department of Physics, Massachusetts Institute of Technology, Cambridge, MA 02138, USA}
\author{Brian J. LeRoy}
\email{leroy@physics.arizona.edu}
\affiliation{Physics Department, University of Arizona, 1118 E 4th Street, Tucson, AZ 85721, USA}
\date{\today}

\begin{abstract}
A perpendicular electric field breaks the layer symmetry of Bernal-stacked bilayer graphene, resulting in the opening of a band gap and a modification of the effective mass of the charge carriers.  Using scanning tunneling microscopy and spectroscopy, we examine standing waves in the local density of states of bilayer graphene formed by scattering from a bilayer/trilayer boundary.  The quasiparticle interference properties are controlled by the bilayer graphene band structure, allowing a direct local probe of the evolution of the band structure of bilayer graphene as a function of electric field.  We extract the Slonczewski-Weiss-McClure model tight binding parameters as $\gamma_0 = 3.1$ eV, $\gamma_1 = 0.39$ eV, and $\gamma_4 = 0.22$ eV.  
\end{abstract}

\maketitle
Quasiparticle interference in metals results in standing waves in the local density of states (LDOS) whose properties reflect both the nature of the defect scatterer and the underlying electronic properties of the material itself.  For the case of scattering from a straight edge, these Friedel oscillations have been studied at length for metals such as Cu ~\cite{Crommie1993} and Au ~\cite{Hasegawa1993}.  They have recently been examined in monolayer graphene as well, where a step in a hexagonal boron nitride (hBN) substrate acted as the scatterer ~\cite{Xue2012}.  These measurements allow direct reconstruction of the Fermi surface properties of the material carrying the LDOS waves.  Peculiar properties of monolayer graphene, such as its small Fermi surface at low energy and the pseudospin of its charge carriers, result in Friedel oscillations with unusually long wavelength and faster decay than those seen in noble metals.  The Friedel oscillations in Bernal-stacked bilayer graphene are expected to be of similarly unusual nature as those observed in monolayer graphene.  Differences in the exact properties of these Friedel oscillations should arise due to the hyperbolic band structure of bilayer graphene, as opposed to the linear band structure of monolayer graphene ~\cite{Rutter2007,Brihuega2008,Mallet2012}.  The ability to tune the band gap of bilayer graphene with an electric field ~\cite{McCann2006} adds an additional degree of freedom to these Friedel oscillations, whereas the band structures of metals and monolayer graphene are independent of electric field.

In this letter, we present scanning tunneling microscopy (STM) and spectroscopy (STS) measurements of exfoliated few-layer graphene on hBN.  Specifically, we examine interfaces between bilayer and trilayer graphene, where the layer-change region acts as an edge scatterer for carriers in the bilayer graphene.  By tracking the behavior of Friedel oscillations in bilayer graphene, we are able to map its band structure as a function of electric field.  Prior measurements have indirectly mapped the band structure of bilayer graphene via optical spectroscopy and transport measurement techniques ~\cite{Ohta2006, Malard2007, Zhang2008, Li2009, Kuzmenko2009a, Kuzmenko2009b, Zou2011,Lee2014}.  In the former case, features of the band structure were deduced from optical transition energies, and in the latter from cyclotron mass and chemical potential measurements.  The band structure of bilayer graphene has also been examined locally with STM ~\cite{Rutter2007,Mallet2012}, but without the ability to probe both bands or tune the electric field.  This work provides the first direct probe of the full energy versus momentum dispersion of bilayer graphene as a function of electric field.

\begin{figure}[t]
\includegraphics[width=8.5cm]{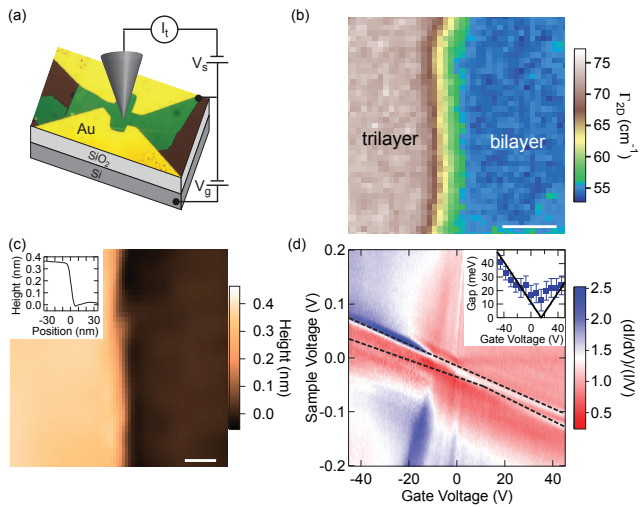} 
\caption{Experimental setup. (a) Schematic of the measurement setup showing the STM tip and an optical microscope image of a measured sample of few layer graphene on hBN. (b) Raman spectroscopy mapping of the bilayer/trilayer interface. Scale bar is 500 nm. (c) STM topography across a trilayer/bilayer (left/right) interface showing a sharp layer change transition. Scale bar is 10 nm.  Inset: averaged cut of the topography showing a step height of roughly 0.35 nm. (d) Normalized (dI/dV)/(I/V) spectroscopy on bilayer graphene as a function of gate voltage.  Black dotted lines are guides to the eye marking the approximate valence and conduction band edges. Inset: experimentally extracted gap size as a function of gate voltage in blue squares from maps similar to those in Fig. ~\ref{fig:cuts}. Black line is the gap assumed for the band structure fitting in Fig. ~\ref{fig:dispersion}.  Error bars represent the uncertainty in determining the band edge.}
\label{fig:schematic}
\end{figure}

Fig. ~\ref{fig:schematic}(a) shows a schematic diagram of the measurement set-up used for STM imaging and spectroscopy of the few-layer graphene flakes.  All the STM measurements were performed in ultrahigh vacuum at a temperature of 4.5 K.  dI/dV spectroscopy measurements were acquired by turning off the feedback circuit and adding a small (5 mV) a.c. voltage at 563 Hz to the sample voltage.  The current was measured by lock-in detection.  Fig. ~\ref{fig:schematic}(b) shows Raman spectroscopy mapping of a representative graphene on hBN flake measured in this study.  A region of bilayer graphene neighbors trilayer graphene in a continuous sheet.  These regions are identified by a change in the width of the Raman 2D peak ~\cite{Malard2009}.

Fig. ~\ref{fig:schematic}(c) shows STM topography of the bilayer/trilayer interface, and the inset shows an averaged cut of the height.  From this, we find that the edge appears as a clean step height of approximately 0.35 nm, which is consistent with a one additional layer of graphene ~\cite{CastroNeto2009}.  We do not observe a sizable moir\'e pattern in any of the samples presented, eliminating the need to consider any modification of the density of states at low energy due to substrate interactions ~\cite{Yankowitz2012}.  Fig. ~\ref{fig:schematic}(d) shows normalized (dI/dV)/(I/V) spectroscopy as a function of gate voltage taken in the bilayer region, far from the trilayer boundary.  The approximate valence and conduction band edges are marked with black dotted lines, indicating a field tunable band gap as large as about 45 meV at our most negative gate voltages probed.  The electric field arises in a small region underneath the tip due to the voltage differences between the STM tip, silicon back gate and graphene. 

\begin{figure}[t]
\includegraphics[width=13.65cm]{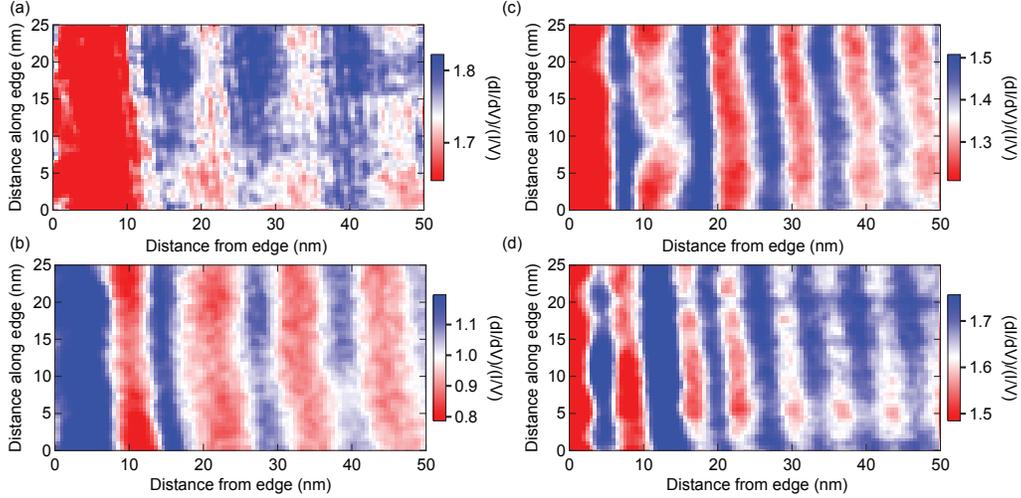} 
\caption{Standing waves in the LDOS of bilayer graphene near a trilayer graphene edge.  Maps show normalized dI/dV spectroscopy and are taken at sample voltages of (a) +130 mV (b) +10 mV (c) -40 mV and (d) -90 mV (the local charge neutrality point is at about +45 mV).  (a) probes the conduction band (electrons) and (b) - (d) probe the valence band (holes).  The data is acquired at $V_g$ = -60 V.}
\label{fig:waves}
\end{figure}

Figures ~\ref{fig:waves}(a) - (d) show maps of normalized dI/dV spectroscopy acquired within a few nanometers of the bilayer/trilayer edge interface.  All four maps are taken at the same spatial location and back gate voltage, and differ only in sample voltage.  The maps are taken completely on the bilayer graphene region of the sample, with the trilayer graphene edge just outside the left-hand side of each image.  We find spatially-coherent long-wavelength standing waves in the density of states which emanate roughly parallel to the edge.  The wavelength of these oscillations depends on the sample voltage applied.  The maps presented here were taken at a back gate voltage of $V_g$ = -60 V, although similar standing waves are observed at all back gate voltages probed.  Standing waves can generally be observed on both the electron and hole sides of the charge neutrality point (CNP), although smearing at higher sample voltages tends to attenuate the waves from one of the two bands, especially if the band edges are far from the Fermi energy.  The map in Fig. ~\ref{fig:waves}(a) probes the conduction band (electrons), while Figs. ~\ref{fig:waves}(b) - (d) probe the valence band (holes).  We have measured two similar samples (one neighboring ABA- and the other ABC-stacked trilayer graphene) and observed similar standing waves in both.  The stacking configuration of the trilayer graphene does not play a role in the wavelength of the LDOS oscillations in bilayer graphene and therefore we will concentrate on the results from just one stacking configuration.

\begin{figure}[t]
\includegraphics[width=8cm]{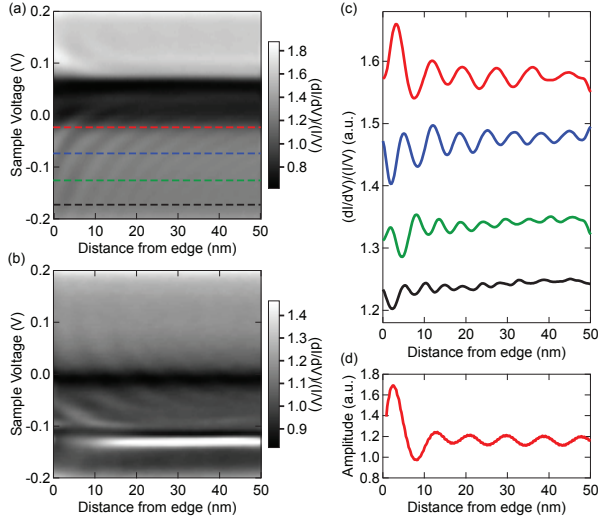} 
\caption{Dispersion of LDOS waves with energy.  Normalized dI/dV maps as a function of sample voltage and distance from the edge at back gate voltages of (a) $V_g$ = -45 V and (b) $V_g$ = +45 V. (c) Constant sample voltage cuts of (a) as marked by the color-coded dashed lines.  The curves are taken at sample voltages of -25 mV, -75 mV, -125 mV, and -175 mV, from top to bottom.  Curves are offset for clarity. (d) DFT simulation of the LDOS in bilayer graphene neighboring an ABC-stacked trilayer graphene edge with zigzag termination at E = +100 meV in zero electric field.}
\newpage
\label{fig:cuts}
\end{figure}

Comparable data sets to those presented in Figs. ~\ref{fig:waves}(a) - (d) can be vertically averaged (along the direction of the edge) to present the standing waves as a function of sample voltage in a single plot, as is shown in Figs. ~\ref{fig:cuts}(a) and (b) for back gate voltages of -45 V and +45 V, respectively.  This gives a visual of the dispersion of the standing waves as a function of both sample voltage and distance of the STM tip from the bilayer/trilayer interface.  In addition to the standing waves, horizontal regions of low dI/dV spectroscopy are present in both Figs. ~\ref{fig:cuts}(a) and (b), centered around a sample voltages of about +50 mV and -110 mV, respectively.  These regions represent the band gap of bilayer graphene and do not exhibit standing waves (consistent with a gapped system).  There is an additional horizontal region of low normalized dI/dV spectroscopy in both Figs. ~\ref{fig:cuts}(a) and (b) at the Fermi energy (0 V in sample voltage), which is due to the decreased tunneling probability at zero bias on the STM tip and is not related to a feature of the bilayer graphene band structure.  Thus, data very close to the Fermi energy is ignored for this study.

\begin{figure}[t]
\includegraphics[width=6cm]{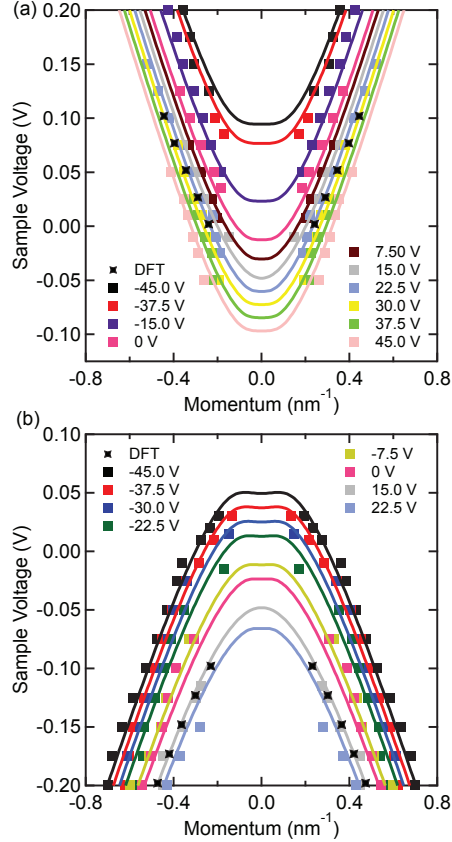} 
\caption{Experimentally extracted energy versus momentum as a function gate voltage, and theoretical fits. (a) Colored squares represent the fits of the momentum as a function of energy extracted from the wavelength of the LDOS standing waves for the conduction band (electrons).  The solid line is the theoretical best fit of the dispersion. (b) Same as (a) for the valence band (holes). Best-fit lines are in the same color coding as the experimental points for each gate voltage.  The gate voltage of each measurement is indicated in the color key.  The black star symbols in both (a) and (b) represent the results from DFT calculations in zero electric field, which roughly corresponds to $V_g$ = +15 V.}
\newpage
\label{fig:dispersion}
\end{figure}

Figure ~\ref{fig:cuts}(c) shows cuts of Fig. ~\ref{fig:cuts}(a) at constant sample voltages.  We can fit the individual standing wave cuts using a function proportional to $\cos{(2kx)}$, where $k$ is the momentum of the charge carriers and $x$ is the distance from the edge ~\cite{Xue2012}.  Due to the complication of different pseudospins in the bilayer and trilayer regions and the unknown atomic edge configuration terminating the trilayer edge, we are unable to calculate the exact amplitude decay factor.  However, the wavelength of the standing waves is independent of the choice of this decay, so a power law decay of the form $1/x$ was chosen as it roughly matched the majority of the data.  We fit the wavelength of the standing waves over a 400 meV range of sample voltages and plot the extracted momentum as the colored squares in Figs. ~\ref{fig:dispersion}(a) and (b).  The color corresponds to the gate voltage at which the data was acquired, with gate voltages ranging from $V_g$ = +45 V to -45 V in steps of 7.5 V.  Fig. ~\ref{fig:dispersion}(a) shows the results of the electron fits (conduction band), while Fig. ~\ref{fig:dispersion}(b) shows the results of the hole fits (valence band).  Depending on the strength of the standing waves, fits could be done for both bands for some gate voltage maps, while others only exhibited strong enough waves for one of the two bands to be fit.  The bands move in sample voltage for different gate voltages primarily due to the shift of the Fermi energy induced by changing the back gate voltage.

To confirm our results, we have simulated the LDOS in bilayer graphene resulting from electron scattering off a trilayer graphene edge.  We performed the simulation using density functional theory (DFT) implemented in ATK~\cite{Soler2002}, with the density matrix calculated using non-equilibrium Green's functions.  As shown in Fig.~\ref{fig:cuts}(d) for E = +100 meV, the simulation also exhibits a standing wave in the LDOS of bilayer graphene ~\cite{phase_comment}.  We chose the ABC stacking for the trilayer graphene to match our data in Fig. ~\ref{fig:dispersion}, although we find virtually no quantitative difference in the LDOS wavelength between the two stacking configurations of trilayer graphene ~\cite{edge_comment}.  We performed the simulation at varying energies, and the resulting energy versus momentum dispersion is plotted in the black star symbols of Fig. ~\ref{fig:dispersion}.  We note that this data should be compared with the experimental data points at $V_g$ = +15 V due to an experimental offset in the zero of the electric field.  We find relatively good quantitative agreement between our experimental data points and the DFT simulations, especially in the valence band.  We see less electron-hole asymmetry in our simulations than in our experimental data, and as a result the simulated conduction band is slightly wider than our experimental results.  However, the overall agreement between the two suggests that we can safely describe the bilayer/trilayer interface scattering (which is complicated by different pseudospins in each layer and an unknown atomic edge configuration) with a simple $2k$ scattering model ~\cite{Xue2012}.   

To analyze the dispersion in Fig. ~\ref{fig:dispersion}, we take a simple tight-binding model Hamiltonian for bilayer graphene
\begin{equation} \label{eq:Hamiltonian}
{H} = \left(
\begin{array}{cccc}
-U/2 & v {\pi}^\dag & -v_4 {\pi}^\dag & 0\\
v \pi & -U/2+\Delta & \gamma_1 & -v_4 {\pi}^\dag\\
-v_4 \pi & \gamma_1 & U/2+\Delta & v {\pi}^\dag\\
0 & -v_4 \pi & v \pi & U/2\\
\end{array}
\right) \, ,
\end{equation}
where $\pi = p_x + i p_y$, ${\pi}^{\dag} = p_x - i p_y$, with $p$ the carrier momentum, $v = \sqrt{3} a \gamma_0 / 2 \hbar$ the Fermi velocity, $v_4 = \sqrt{3} a \gamma_4 / 2 \hbar$ an effective velocity, $U$ the interlayer bias, and $a = 2.46$ {\AA} the graphene lattice constant.  Adopting the atom labeling convention of Ref. ~\onlinecite{McCann2013}, $\gamma_0$ is the nearest-neighbor in-plane hopping energy (A1-B1), $\gamma_1$ is the hopping energy between orbital pairs on the dimer sites (B1-A2), $\gamma_4$ is the hopping energy between dimer and non-dimer orbitals (A1-A2 or B1-B2), and $\Delta$ is the stacking energy difference between dimer and non-dimer sites.  $\gamma_4$ and $\Delta$ control the electron-hole asymmetry (increasing either results in heavier holes and lighter electrons), which we find to be quite prevalent in our measurements.

We additionally account for the change in both the K-point gap size $U$ and the shift of the Fermi energy as a function of the back gate voltage.  By tracking the movement of the band gap center as a function of gate voltage from the individual gate voltage maps (similar to those shown in Figs.~\ref{fig:cuts}(a) and (b)), we find the Fermi energy shifts in sample voltage by -2 mV per volt on the back gate, and is centered about zero sample voltage at $V_g$ = -9 V.  This movement is in reasonable agreement with prior STM measurements of bilayer graphene ~\cite{Deshpande2009,Rutter2011}.  Similarly, by tracking the approximate band edges we extract the band gap (Fig. ~\ref{fig:schematic}(d) inset) as a function of gate voltage.  The direct gap $\tilde{U} = \frac{U}{\sqrt{1+(U/\gamma_{\rm 1})^2}}$ differs from the K-point gap by less than 1\% at our maximum gap values, therefore we ignore this distinction.  We find the gap grows at a rate of 0.75 meV per volt on the back gate, with the minimum gap at $V_g$ = +15 V.  Assuming a model for the STM tip and graphene/back gate system as in Ref. ~\onlinecite{Yankowitz2014}, this gap size as a function of gate voltage is well matched to theoretical calculations for bilayer graphene with screening ~\cite{Zhang2010}.  The difference between the gate voltage of the CNP and band gap minimum is likely due to residual doping on the sample and a work function mismatch between the STM tip and the bilayer graphene.

We fit the electron and hole data at all electric fields simultaneously following Eq. ~\ref{eq:Hamiltonian}, taking $\gamma_1$ = 0.39 eV and $\Delta$ = 0.018 eV as known from prior measurements ~\cite{Ohta2006,Zhang2008,Li2009,Kuzmenko2009a,Kuzmenko2009b,Zou2011}, and leaving $\gamma_0$ and $\gamma_4$ as fitting parameters.  The best fit ~\cite{fit_comment} to the data gives $\gamma_0$ = 3.1 eV (which corresponds to $v_F = 1.0 \times 10^6$ m/s) and $\gamma_4$ = 0.22 eV.  This value of $\gamma_4$ falls just above the high end of prior results obtained via optical spectroscopy ~\cite{Malard2007,Zhang2008,Li2009,Kuzmenko2009b} and magnetoresistance measurements ~\cite{Zou2011}, and represents pronounced electron-hole asymmetry.  The resulting bands are plotted on top of the experimental data points in Fig. ~\ref{fig:dispersion}, and are well fit to the data for all electric fields.

In conclusion, we have examined quasiparticle scattering in bilayer graphene and observe coherent standing waves in the local density of states.  By concurrently tracking the band gap and Fermi surface profile extracted from these standing waves, we can construct a complete picture of the bilayer graphene band structure and its evolution with electric field.  We observe pronounced electron-hole asymmetry in bilayer graphene, which can be captured by the tight-binding model with a large dimer to non-dimer hopping parameter $\gamma_4$.  This work represents the first direct local mapping of the band structure of bilayer graphene as a function of electric field.\\

M.Y. and B.J.L. were supported by the U. S. Army Research Laboratory and the U. S. Army Research Office under contract/grant number W911NF-09-1-0333 and the National Science Foundation CAREER award DMR-0953784.  J.I-J.W. was partially supported by a Taiwan Merit Scholarship TMS-094-1-A-001.  J.I-J.W and P.J-H. have been primarily supported by the US DOE, BES Office, Division of Materials Sciences and Engineering under Award DE-SC0001819. Early fabrication feasibility studies were supported by NSF Career Award No. DMR-0845287 and the ONR GATE MURI. This work made use of the MRSEC Shared Experimental Facilities supported by NSF under award No. DMR-0819762 and of Harvard's CNS, supported by NSF under grant No. ECS-0335765.  A.G.B. was supported by the U.S. Army Research Laboratory (ARL) Director's Strategic Initiative program on interfaces in stacked 2D atomic layered materials.  S.L. is supported by the A*STAR Graduate Scholarship. S.Y.Q. is supported by the IHPC Independent Investigatorship and Singapore NRF Fellowship (NRF-NRFF2013-07).  We thank NUS Graphene Research Centre, A*CRC, and Prof. Feng Yuan Ping's Lab in NUS for computational support.

\newpage

\begin{thebibliography}{99}
\bibitem{phase_comment} The A and B graphene sublattices each contribute a slightly different phase to the standing wave. These two phases overlap significantly 10 nm and further from the interface, and only show small divergence close to the edge.  Since both phases exhibit nearly equivalent overlapping wavelengths, we only plot the phase component from the A-sublattice for clarity.
\bibitem{edge_comment} We chose a zigzag edge termination for the trilayer region, but do not find the specific edge termination to significantly affect the wavelength either.
\bibitem{fit_comment} There is no simple algebraic expression for the dispersion when including $\gamma_4$.  Therefore, our best-fit is determined by hand.  However, the results are quite sensitive to the free fitting parameters $\gamma_0$ and $\gamma_4$, therefore they are reasonably precise.
\end{thebibliography}
\end{document}